# Additive noise induces Front propagation

M.G. Clerc, C. Falcon and E.Tirapegui
*Departamento de Física, Facultad de Ciencias Físicas y Matemáticas,Universidad de Chile, Casilla 487-3, Santiago, Chile.*

The effect of noise in a motionless front between a periodic spatial state and an homogeneous one is studied. Numerical simulations show that noise induces front propagation. From the subcritical Swift-Hohenberg equation with noise, we deduce an adequate equation for the envelope and the core of the front. The equation of the front's core is characterized by an asymmetrical periodic potential plus additive noise. The conversion of random fluctuations into direct motion of front core is responsible of the propagation. We obtain an analytical expression for the velocity of the front, which is in good agreement with numerical simulations.

The description of macroscopic matter, i.e. matter composed by a large number of microscopic constituents, is usually done using a small number of coarse-grained or macroscopic variables. When spatial inhomogeneities are considered these variables are spatio-temporal fields whose evolution is determined by deterministic partial differential equations (PDE). This reduction is possible due to a separation of time scales, which allows a description in terms of the slowly varying macroscopic variables, which are in fact fluctuating variables due to the elimination of a large number of fast variables whose effect can be modelized including suitable stochastic terms (noise) in the PDE. The influence of noise in nonlinear systems has been the subject of intense experimental and theoretical investigations [1–8]. Far from being merely a perturbation to the idealized deterministic evolution or an undesirable source of randomness and disorganization, noise can induce specific and even counterintuitive dynamical behavior. The most well-know examples in zero dimensional systems are noise induced transition [1] and stochastic resonance [2]. More recently, examples in spatial extended system are noise induced phase transition [3], noise-induced patterns [4–6], stochastic spatio-temporal intermittency [7] and noise-induced travelling waves [8]. Here, we will focus on the effect of additive noise in front propagation. The concept of front propagation emerged in the field of populations dynamics [9], and the interest in this type of problems has been growing steadily in Chemistry, Physics and Mathematics. In Physics, front propagation plays a central role in a large variety of situations, ranging from reaction diffusion models to general pattern forming systems (see the review [10] and references therein). The influence of multiplicative noise in a globally stable state invading an unstable or metastable state, a *front solution*, has been extensively studied in the literature, particularly concerning the issue of velocity selection[11].

The aim of this article is to study the effect of noise in a motionless front between a periodic spatial state and an homogeneous state, both stable. Numerical simulation of this type of front shows that noise induces front propagation, that is, one state invades the other one. From a prototype model that exhibits this type of front, the subcritical Swift-Hohenberg equation with additive noise, we deduce an equation for the envelope of the front solution, which includes non resonant terms. From this equation we derive an equation for the core of the front, which is characterized by a periodic asymmetrical potential with stable equilibria plus additive noise. The conversion of random fluctuations into directed motion of the core of the front is responsible of front propagation. We obtain an analytical expression for the velocity of the front, which is proportional to Kramer's rate, in the weak noise intensity limit. This expression is in good agreement with numerical simulations.

A front between two stable homogeneous states in a variational system with a known free energy, permanently propagates from the state with higher free energy to the state with lowest free energy [10]. However, the front is static when both states are energetically equal, i.e. in the *Maxwell point*. This picture changes in the case of a front between a periodic spatial state and an homogeneous one. The front exhibits a *locking phenomena* in a region of parameters known as the pinning range [12], in which the front does not move. When additive white noise is taking into account, one may expect ran-dom fluctuations of the interface between the two states (core of the front). However, numerical simulations in a one dimensional extended system show that the front propagates from one state to the other with a well defined average velocity, as it is illustrated in Fig. 1. Depending on the region of parameters, the front can propagate to the periodic spatial state or to the homogeneous state.

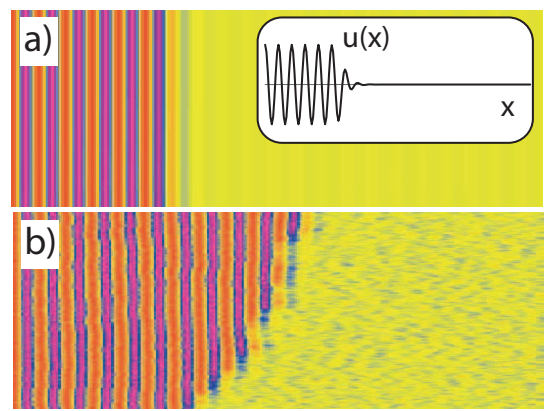

FIG. 1: Spatiotemporal evolution of Eq. (1), with time running up. The gray scale is proportional to field $u$. The inset figure is the initial condition. The parameters have been chosen $\epsilon = -0.16$, $\nu = 1.0$, $q = 0.7$ a) $\eta = 0.0$, and b) $\eta = 0.4$

In order to understand the mechanism through which noise induces propagation, we consider a prototype model that exhibits this type of front (subcritical Swift-Hohenberg equation with noise [10])

$$\partial_t u = \varepsilon u + \nu u^3 - u^5 - (\partial_{xx} + q^2)^2 u + \sqrt{\eta}\zeta(x,t), \quad (1)$$

where $u(x,t)$ is an order parameter, $\varepsilon - q^4$ is the bifurcation parameter, $q$ is the wave-number of periodic spatial solutions, $\nu$ the control parameter of the type of bifurcation (supercritical or subcritical), $\zeta(x,t)$ is a gaussian white noise with zero mean value and correlation $\langle\zeta(x,t)\zeta(x',t')\rangle = \delta(x-x')\delta(t-t')$, and $\eta$ represents the intensity of the noise. The model (1) describes the confluence of a stationary and an spatial subcritical bifurcation, when the parameters scale as $u \sim \varepsilon^{1/4}$, $\nu \sim \varepsilon^{1/2}$, $q \sim \varepsilon^{1/4}$, $\partial_t \sim \varepsilon$ and $\partial_x \sim \varepsilon^{1/4}$ ($\varepsilon \ll 1$). This bifurcation is of codimension three. The above model is often employed in the description of patterns observed in Rayleigh-Benard convection [10].

For small and negative $\nu$ and $9\nu^2/40 < \varepsilon < 0$, the system exhibits coexistence between a stable homogenous state $u = 0$ and a periodic spatial one $u = \sqrt{\nu}\left(\sqrt{2(1+\sqrt{1+40\varepsilon/9\nu})}\cos(qx)\right) + o(\nu^{5/2})$. In this parameter region, one finds a front between these two states. These type of solution is an heteroclinic curve of the spatial dynamical system associated to the above model [13]. A front between a homogeneous and a spatial oscillatory state can be described by the ansatz

$$u = \sqrt{\frac{2\nu}{10}}\varepsilon^{1/4}\left\{A\left(y = \frac{3\sqrt{|\varepsilon|}}{2\sqrt{10}q}x, \tau = \frac{9\nu^2|\varepsilon|}{10}t\right)e^{iqx}\right.$$
$$\left.+w_1(x,y,\tau) + c.c\right\}, \quad (2)$$

where $A(y,\tau)$ is the envelope that describes the front solution, $w_1(x,y,\tau)$ is a small correction function of order $\varepsilon$, and $\{y,\tau\}$ are slow variables. In this ansatz (2), we consider $q$ is order one or larger that the other parameters. Introducing the above ansatz in equation (1) and linearizing in $w_1$, we find the following solvability condition

$$\partial_\tau A = \epsilon A + |A|^2 A - |A|^4 A + \partial_{yy}A$$
$$+\left(\frac{A^3}{9\nu} - \frac{A^3|A|^2}{2}\right)e^{\frac{2iqy}{a\sqrt{|\varepsilon|}}} - \frac{A^5}{10}e^{\frac{4iqy}{a\sqrt{|\varepsilon|}}} + \frac{\sqrt{\eta}b}{|\varepsilon|^2}e^{\frac{iqy}{a\sqrt{|\varepsilon|}}}\zeta(y,\tau),$$
$$(3)$$

where $\epsilon \equiv 10\varepsilon/9\nu^2$, $a \equiv 3\nu/2\sqrt{10}q$, and $b \equiv 10^{9/4}/81\nu^4$. The deterministic terms proportional to the exponential are non resonant, that is, one can eliminate these terms by an asymptotic change of variable. Furthermore, they have rapidly varying oscillations in the limit $\epsilon \to 0$. Hence, one usually neglects these terms. When one considers only the deterministic resonant terms, first line of (3), it is straightforward to show that the system exhibits a front solution between two homogeneous states, 0 and $(1+\sqrt{1+4\epsilon})/2$, when $\epsilon < 0$. This front propagates from the stable state (lowest free energy) to the metastable one, and it is static when $\epsilon_M = -3/16$, i.e. in the *Maxwell point,* and it has the form

$$A_\pm = \sqrt{\frac{3/4}{1+e^{\pm\sqrt{3/4}(y-y_o)}}}e^{i\theta},$$

where $y_o$ is the position of the front's core, and $\theta$ is an arbitrary phase. In the neighborhood of $\epsilon_M$ the front propagates by a velocity given approximately by $\Delta = 3d/8(\epsilon - \epsilon_M)$, where $d = \int(\partial_y A_+)^2 dy$. However, as pointed out by Pomeau [12], static fronts between a homogeneous and spatial periodic state may actually persist in a finite neighborhood of the Maxwell point, the *pinning range,* and it was conjectured that this phenomena could be due to non adiabatic effects produced by non resonant terms. This was shown in a particular case in [14] and has recently been discussed in a general frame in [15], with the conclusion that the locking phenomena results from the interaction (contained in the non resonant terms) of the large scale envelope $A(y,\tau)$ with the small scale underlying the spatial periodic solution [14].

In order to describe the dynamics exhibited by (1) and the locking phenomena, we must then consider the non resonant terms in the envelope equation (3). We consider all these terms as perturbations because they have rapidly varying oscillations. Close to the Maxwell point, we use the ansatz

$$A(y,\tau) = (A_+(y-y_o(\tau)) + \delta\rho)e^{i\delta\Theta}$$

where $\delta$ is a small parameter order $(\epsilon - \epsilon_M)$. Introducing the above ansatz in equation (3) and linearizing in $\{\rho,\Theta\}$

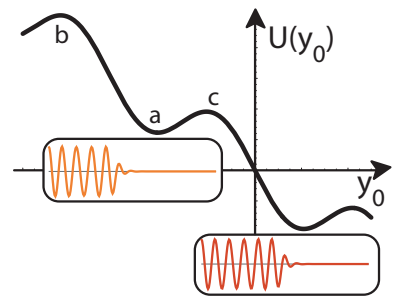

FIG. 2: Schematic representation of the potential $U(y_0)$ of Eq. (4). $\{a,b,c\}$ are fixed points. The inset figure represent two successive equilibria state of Eq. (1).

we obtain the following solvability condition

$$\dot{y}_o = -\frac{\partial U(y_0)}{\partial y_o} + \frac{ab}{|\epsilon|^2}\sqrt{\frac{\eta}{2d}}\zeta(\tau)$$
$$= \Delta + \Gamma\cos\left(\frac{2q}{d\sqrt{|\epsilon|}}y_o - \varphi\right) + \frac{ab}{|\epsilon|^2}\sqrt{\frac{\eta}{2d}}\zeta(\tau) \quad (4)$$

where $\Gamma \equiv \sqrt{k_1 + k_2}e^{-\sqrt{4/3}\pi q/d\sqrt{|\epsilon|}}$, $\tan\varphi = k_1/k_2$, $k_1 \equiv -9d\pi/2048\ (8q/a\sqrt{3|\epsilon|})^3 - (8+d\pi/32\nu)(8q/a\sqrt{3|\epsilon|}) + \sqrt{3}d^2\beta^2q^3\eta\pi/2^3a^3|\epsilon|^{11/2}$, and $k_2 \equiv (27/1024 - 1/128\nu)d\pi\ (8q/a\sqrt{3|\epsilon|})^2 - 3d^2\beta^2q^2\eta\pi/2^6a^2|\epsilon|^5$. $U(y_0)$ is the potential which characterizes the dynamics of the front's core and $\zeta(\tau)$ is a gaussian white noise, that is, with zero mean value and correlation $\langle\zeta(\tau)\zeta(\tau')\rangle = \delta(\tau-\tau')$.

Due to the interaction of the large scale with the small scale underlying the spatial periodic solution, the dynamics of the front's core is modified with terms which are exponentially small and periodic in space. The above deterministic system is characterized by the spatial periodic state invading the homogeneous one with a well defined velocity when $\Delta < 0$ and $|\Delta| > |\Gamma|$. Increasing $\Delta$, the system exhibits a simultaneous transition to infinite saddle nodes for $|\Delta_-| = |\Gamma|$. For $\Delta > \Delta_-$ and $|\Delta| < |\Gamma|$, the system has an infinite number of stable equilibria. Each equilibrium point represents a static front with different bumps (see Fig. 2). Increasing further $\Delta$, all critical points disappear by saddle-node when $\Delta > 0$ and $|\Delta_+| = |\Gamma|$. For $\Delta > \Delta_+$ the homogeneous states invades the spatial periodic one with a well defined velocity. Therefore, for $\Delta_- < \Delta < \Delta_+$ (pinning range) the system exhibit the locking phenomena. We now consider the effect of noise in (4). Due to the asymmetry of the potential the system does not have a global stationary state and continuously converts the random fluctuations in directed motion of the front, i.e. the noise induces front propagation. This type of phenomena is well known as a Brownian motor [16]. One can easily understand the origin of this phenomena: if initially $y_o$ is inside the basin of attraction $U$ of a fixed point the front just fluctuates around the fixed point during a time of the order of the mean first passage time to $\partial U$, the border of $U$, after this time the system makes a transition to the basin of attraction of the nearest stable fixed point separated from the first one by the lowest energy barrier. This behavior is repeated in this new basin of attraction and the final result is a directed motion of the front. Since the energy threshold for jumping to the right or to the left are different the probability of jumping to the side with the highest energy threshold will be exponentially small with respect to the probability of jumping to the other side and this determines the direction of motion of the front.

From the above analysis, we can estimate the mean velocity of the front's core

$$\langle v \rangle = \frac{\pi\sqrt{|\epsilon|}}{qa}\left(\frac{1}{\tau_+} - \frac{1}{\tau_-}\right)$$

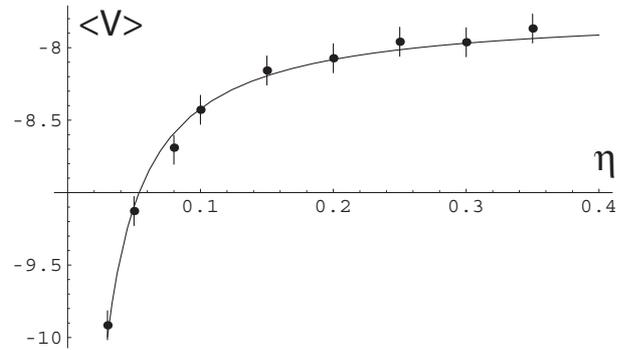

FIG. 3: Mean velocity of the front. The continuous lines is the analytical formula of mean velocity and the dot points are numerical measuring of the mean velocity of the front of Eq. (1).

where $\pi\sqrt{|\epsilon|}/qa$ is the distance between the two successive fixed points and $\{\tau_-,\tau_+\}$ are the escape times to move to the basins of attraction of the left or right fixed point, respectively. They have the expression

$$\left(\frac{1}{\tau_+} - \frac{1}{\tau_-}\right)^{-1} = \frac{2}{\theta}\int_{c'}^{b'}\int_{c'}^{y}e^{2[U[y]-U[z]]/\theta}dydz$$
$$-\frac{2}{\theta}\int_{c'}^{a'}\int_{c'}^{y}e^{2[U[y]-U[z]]/\theta}dydz\left[\frac{\int_{c'}^{a'}e^{2U[y]/\theta}dy}{\int_{c'}^{b'}e^{2U[y]/\theta}dy}\right] \quad (5)$$

where $a'$, $b'$, and $c'$ are a minimum and two successive maximums of the potential $U(y_0)$ (See Fig. 2), respectively, and $\theta \equiv ab/|\epsilon|^2\sqrt{\eta/2d}$. In the limit of weak noise

$$\langle v \rangle = \frac{2\sqrt{|\epsilon|}}{qa\sqrt{\partial_{yy}U(a')|\partial_{yy}U(c')|}}e^{-\frac{(U(c')-U(a'))}{\theta}}$$
$$\left(1 - \sqrt{\frac{|\partial_{yy}U(c')|}{|\partial_{yy}U(b')|}}e^{-\frac{(U(b')-U(c'))}{\theta}}\right).$$

From the above expression one can find that in this limit the velocity is proportional to Kramer's rate. Numerically, we have measured the front velocity for different values of the noise intensity and we obtain a good agreement with the theoretical prediction, as it is shown in Fig. 3. It is important to remark that $U(y_o)$ is function of the noise intensity. For finite noise intensity this dependence is dominant in the terms $k_1$ and $k_2$, in the limit of $\epsilon \to 0$. Hence for finite noise intensity one only needs to consider the terms coming from the noise to explain the locking phenomena and the induced front propagation.

In order to understand the mechanism of noise induced front propagation we have considered the subcritical Swift-Hohenberg equation. This model allows us to obtain analytical expressions for the mean velocity of the front. For an arbitrary model it is thorny to obtain explicit formulas for the front velocity, since in general we do not have access to explicit expressions of spatial periodic solutions. Given a system that exhibits locking phenomena between a spatial periodic state

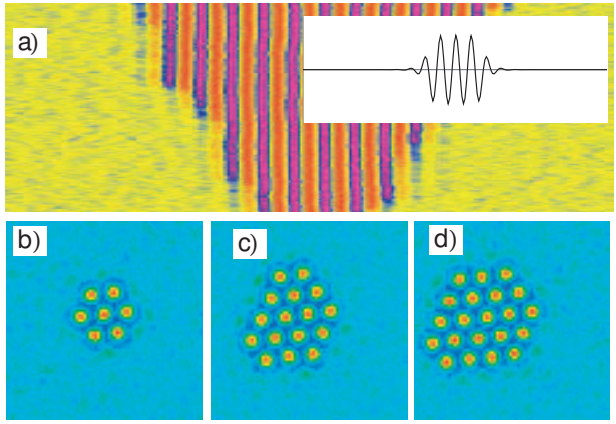

FIG. 4: Noise induces propagation of localized patterns. a) Numerical simulations of generalized Swift-Hohenberg model in one-extended system with additive noise. The inset figure is the initial condition. b), c), and d) are different snapshots of sequent time for numerical simulations of generalized Swift-Hohenberg model in two-extended system with noise

and an homogeneous state, we expect to found, close to a spatial bifurcation, an adequate envelope equation since the spatial periodic solutions in the onset of the bifurcation are harmonic and coexists with the homogeneous state. Hence, one can use an ansatz similar to (2) and noticing that the envelope satisfies the symmetries $\{x \to -x, \ A \to \bar{A}\}$, and $\{x \to x+x_o, \ A \to Ae^{iqx_o}\}$[15], we can conclude that the amplitude equation has a form similar to (3) with real coefficients which can be written in the form

$$\partial_T A = f\left(|A|^2\right) A + \partial_{XX} A + \sum_{m,n} g_{mn} A^m \bar{A}^n e^{iq(1+n-m)x}$$

where the terms which have explicit exponential are non resonant and rapidly varying in space. However it is precisely due to these terms that we can explain the locking phenomena. One obtains similar terms (3), which can be dominant, when noise is considered. We can conclude then that the noise induces front propagation due to the asymmetry of the front's core potential and the lack of a global stationary state. Another way to understand this phenomenon is: noise prefers to create or remove a bump, because the necessary perturbations to nucleate or destroy a bump are different.

The existence, stability properties, and bifurcation diagrams of localized patterns in the pinning range in one dimensional extended systems have recently been studied [13], from a dynamical point of view. When we consider the effects of noise on these solutions, we expect, due to our previous discussion, propagation of the interface of these localized patterns. In Fig. 4 we show, in one and two extended dimensions, the noise induced propagation of one state into the other. In one spatial dimension, one can then understand the localized pattern solutions as the interaction of two fronts. In two dimensional spatial systems, the understanding of the phenomena is in progress. From the above results, one realizes that the localized patterns are unstable in nature, that is, in presence of noise. The velocity of propagation of the interfaces and fronts are proportional to Kramer's rate. Therefore, experimentally, one can observe these localized patterns, when noise is weak enough, for long intervals of time, as metastable states.

The simulation software *DimX* developed by P. Coullet and collaborators at the laboratory INLN in France has been used for all the numerical simulations. M.G.C. and E.T. acknowledge the support of FONDECYT projects 1020782 and 1020374, FONDAP grant 11980002, and ECOS-CONICYT collaboration programs.